\documentclass[twocolumn,showpacs,prl,aps,superscriptaddress]{revtex4-1}
\usepackage{ulem,bm}
\usepackage{amsmath, upgreek, amssymb, graphicx, paralist, gensymb,epstopdf}
\usepackage[colorlinks, linkcolor=blue, citecolor=blue, urlcolor=blue, breaklinks=true]{hyperref}
\usepackage{array,epsfig,amsmath,amssymb}
\usepackage{amsmath,epstopdf}
\usepackage{graphicx}
\usepackage{dsfont}
\usepackage{color}
\usepackage{bbm}
\begin{document}
\title{
Coarsening Measurement References and  the Quantum-to-Classical Transition
}

\author{Hyunseok Jeong}
\affiliation{Center for Macroscopic Quantum Control, Department of
Physics and Astronomy, Seoul National University, Seoul, 151-742, Korea}
\author{Youngrong Lim}
\affiliation{Center for Macroscopic Quantum Control, Department of
Physics and Astronomy, Seoul National University, Seoul, 151-742, Korea}
\author{M. S. Kim}
\affiliation{QOLS, Blackett Laboratory, Imperial College London, SW7 2BW, UK}
\date{\today}

\begin{abstract}
We investigate the role of inefficiency in quantum measurements in  the quantum-to-classical transition, and
consistently observe the quantum-to-classical transition by coarsening the 
references of the measurements ({\it e.g.}, when and where to measure).
Our result suggests that the definition of measurement precision in quantum theory 
should include the degree of the observer's ability to precisely control the measurement references.
\end{abstract}


\maketitle

{\it Introduction.-} 
Typical quantum phenomena observed on microscopic scales somehow disappear on macroscopic scales.
There have been trials to explain the quantum-to-classical transition.
Decoherence is one of the well known and successful attempts to explain such a revelation of a classical world
out of quantum mechanical rules \cite{decoherence}.
There are two crucial elements in the framework of quantum mechanics:
one is the {\it state} of a physical system represented by a wave function, and the other is the {\it measurement}
represented by non-negative operators. 
The decoherence program focuses on the evolution of the state: it describes 
a transition of a quantum state to a classical one due to its interactions with environments.

Recently, a different point of view  was presented~\cite{Brukner07},
where coarsening of measurements is attributed to the cause of
the quantum-to-classical transition.
Along this line, it was also pointed out that
coarsening of measurements makes it hard to  detect  micro-macro entanglement in optical systems
\cite{m-m}. 
However, there exist seemingly contradicting results
where even fuzzy measurements allow to observe
severe violations of the Bell inequality~\cite{JeongPRL2009,Lim} and also of
the Leggett-Garg inequality~\cite{Brukner08}. It means that fuzziness in measurements
do not always result in the quantum-to-classical transition.
There is yet another  example in which coarsening measurements results in local realism under stronger restrictions
\cite{Kaszlikowski}. 
There have been extensive attempts to clarify sophisticated conditions of the quantum-to-classical transition  \cite{Buzek1995,JR2006,Cal2011,Kast2013,YHK2013} and it has been found that the quantum-to-classical transition does not always occur when it is expected \cite{JR2006,Kast2013,YHK2013}.
Indeed,  a condition of the measurement process in which the  quantum-to-classical transition is definitely forced to occur is yet to be found.

In fact, a complete measurement process is composed of two parts:
the one is to set a measurement reference and control it while the other is the final detection
with the corresponding projection operator.
The aforementioned works to explain the quantum-to-classical transition
have focused on the role of inefficiency in the final detection
by coarsening its measuring resolution.
On the other hand, the control of the measurement reference is described by 
an appropriate unitary operator with a reference variable applied to the projection operator.
It is worth investigating the role of the measurement reference by
coarsening the accuracy of this unitary operation.
Such unitary operations are often indispensable when strong quantum effects,
incompatible with classical physics, are observed
by standard tools such as Bell's inequality \cite{Bell} and the Leggett-Garg inequality \cite{LeggettGarg}.

Does the accuracy of controlling such measurement 
references play a crucial role in the quantum-to-classical transition?
In this paper, we intensively tackle this question using a generic example of macroscopic entanglement
together with specific physical examples.
Our study clearly shows that  coarsening of the final measurement resolution and that of the measurement reference 
lead to completely different results. 
The quantum-to-classical transition is forced to occur when the
reference of measurement is coarsened, while it is not the case when only the final projection is coarsened.
This aspect of the ``accuracy of the measurement reference'' has not received a proper attention in the context
of the quantum-to-classical transition. 
We believe that our discussion, by clarifying the conditions of the quantum-to-classical transition,
sheds light upon the appearance of a classical world from another angle.

{\it Generic study.-}
We first consider a generic example with an infinite dimensional system
together with an orthonormal basis set 
$\{|o_n\rangle\}$ where $n$ takes integer indexes from the minus to the plus infinities.
Let us consider observable $O^k=O^k_+-O^k_-$ where
\begin{equation}
O^k_+=\sum^\infty_{n=k+1}|o_n\rangle\langle o_n|,~~~O^k_-=\sum^k_{n=-\infty}|o_n\rangle\langle o_n|,
\end{equation}
and $O^k$ represents a ``sharp'' dichotomic measurement with eigenvalues $\pm1$.
A fuzzy version of this dichotomic measurement may be written as
\begin{equation}
O_\delta=\sum_{k=-\infty}^{\infty} P_\delta(k) O^k
\label{fuzzy}
\end{equation}
where
 $P_\delta(k)=\frac{1}{\delta \sqrt{2\pi}} \exp[-\frac{k^2}{2\delta^2}]$ 
is the normalized Gaussian kernel with standard deviation $\delta$.
Here, $\delta$ defines the degree of fuzziness in the measurement, {\it i.e.,} {\it the degree of coarsening
in the final measurement resolution}.
We should assume $\delta>1$ in order to satisfy the normalization condition with the 
discrete version of the Gaussian function, however, this does not affect any essential aspects of our discussions.
We then say that two states are macroscopically distinguishable 
if they can be distinguished with a small error probability using $O_\delta$ with a large value of $\delta$.
For example, states $|o_{n}\rangle$ and $|o_{-n}\rangle$  can be discriminated 
with the error probability of 
\begin{equation}
P_e=1 - {\left[ {\sum\limits_{k =  - \infty }^\infty  {\frac{1}{{\delta \sqrt {2\pi } }}{e^{ - {\textstyle{{{k^2}} \over {2{\delta ^2}}}}}}{\chi _{n - k}}} } \right]^2}
\end{equation}
where $\chi_j$ is 1 for $j>0$ ($-1$ for  $j\leq0$).
Naturally, one can introduce a type of entanglement as follows
\begin{equation}
|M_n\rangle=\frac{1}{\sqrt{2}}(|o_n\rangle|o_{-n}\rangle+|o_{-n}\rangle|o_n\rangle)
\label{eq:me}
\end{equation}
which would become macroscopic entanglement when $n$ is sufficiently large.

We now consider a unitary transform,
$U(\theta)$, that  is the rotation between two states $|o_n\rangle,|o_{-n}\rangle$:
\begin{equation}
\begin{aligned}
&U(\theta)|o_n\rangle=\cos\theta|o_n\rangle+\sin\theta|o_{-n}\rangle,\\
&U(\theta)|o_{-n}\rangle=\sin\theta|o_n\rangle-\cos\theta|o_{-n}\rangle.
\label{unitary}
\end{aligned}
\end{equation}
 If one considers measuring a spin-1/2 system or
polarization of a photon, the unitary operation simply implies a rotation of  the measurement axis.
The coarsened version of the unitary operation applied to the projection operator $O_\delta$ can be described as
\begin{equation}
\begin{aligned}
O_{\delta,\Delta}(\theta_0)= \int &d \theta {P_\Delta }(\theta  - {\theta _0})\left[ {U^\dag(\theta )O_\delta{U }(\theta )} \right]
\end{aligned}
\label{couni}
\end{equation}
where $P_\Delta(\theta-\theta_0)$ is the Gaussian kernel centered around $\theta_0$ with standard deviation $\Delta$. 
In contrast to the value of $\delta$ in Eq.~(\ref{fuzzy}),
 $\Delta$ in Eq.~(\ref{couni}) quantifies 
{\it the degree of coarsening in the measurement reference}.

Now, we study the 
Bell-Clauser-Horne-Shimony-Holt (Bell-CHSH)  inequality \cite{Bell,CHSH} using the entangled state in Eq.~(\ref{eq:me}).
The correlation function is the expectation value of the measurement operators as
\begin{equation}
\begin{aligned}
  {E_{\delta ,\Delta }}({\theta _a},{\theta _b}) = \langle O_{\delta ,\Delta}(\theta_a)
   \otimes O_{\delta ,\Delta}(\theta _b)\rangle_{a,b}
\end{aligned} 
\end{equation}
where the average is taken over entangled state $|M_n\rangle_{ab}$.
Let us first consider that the unitary transform $U(\theta)$ can be perfectly controlled
($\Delta=0$) but the final action of measurement is inaccurate ($\delta>1$).
We then obtain an explicit expression of $E_\delta(\theta_a,\theta_b)$ as
\begin{equation}
\begin{aligned}
E_\delta(\theta_a,\theta_b)=\frac{1}{2}\Big[&
f_\delta(n,\theta_1)f_\delta(-n,\theta_2)+f_\delta(-n,\theta_1)f_\delta(n,\theta_2)\\
&+2g_\delta(n,\theta_1)g_\delta(n,\theta_2)\Big]
\end{aligned}
\end{equation}
where $f_\delta(n,\theta)= \sum_{k=-\infty}^{\infty} P_\delta(k)\big(
\cos^2\theta\chi_{n-k}+\sin^2\theta$ $\chi_{-n-k}
\big)$ and  $g_\delta(n,\theta)=\sin\theta\cos\theta\sum_{k=-\infty}^{\infty} P_\delta(k)
\big(\chi_{n-k}-\chi_{-n-k}\big)$.
The Bell function can be obtained as
\begin{equation}
B=E_\delta(\theta_a,\theta_b)+E_\delta(\theta'_a,\theta_b)+E_\delta(\theta_a,\theta'_b)-E_\delta(\theta'_a,\theta'_b),
\end{equation}
which should satisfy $|B|\leq 2$ by the assumption of local realism \cite{CHSH}.
We plot the numerically optimized Bell function in Fig.~1(a)
for $n$ and $\delta$. Obviously,
an arbitrarily large value of $\delta$ can
be compensated by increasing $n$ in order to observe violation of the Bell-CHSH inequality.

We also consider the case in which 
the unitary transform is coarsened while the efficiency of the final measurement is perfect.
In this case, we set $\Delta$ to be nonzero while $\delta=0$.
The explicit form of the correlation function is then obtained as
\begin{equation}
\begin{aligned}
E_\Delta(\theta_a,\theta_b)= -\int_{-\infty}^\infty\int_{-\infty}^\infty d\phi_ad\phi_b&
P_\Delta(\phi_a-\theta_a) P_\Delta(\phi_b-\theta_b)
\\
&\times \cos[2(\phi_a+\phi_b)].
\label{unicorrel}
\end{aligned}
\end{equation}
Obviously, $E_\Delta(\theta_a,\theta_b)$ is {\it independent} from the value of $n$, {\it i.e.},
macroscopicity of  entanglement.
In Fig.~1(b), it is clear that regardless of the value of $n$, the increase of $\Delta$ 
will totally destroy violation of Bell's inequality.
We have analyzed the Bell-CHSH inequality but 
the Leggett-Garg inequality may be considered in the same way by considering a time-dependent unitary operation
$U(\theta)$ with $\theta=\omega t$.
In what follows, we shall investigate specific physical examples  both for the Bell-CHSH and Leggett-Garg inequalities.

\begin{figure}[t]
\centerline{\scalebox{0.38}{\includegraphics{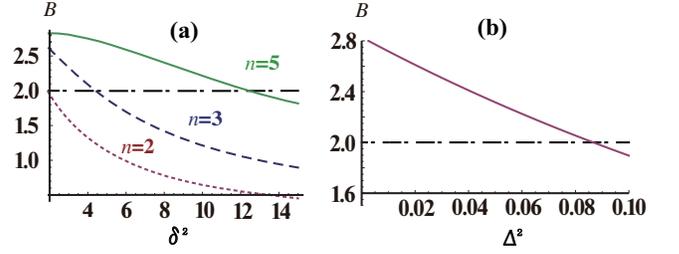}}}
\caption{(color online)
Numerically optimized Bell function $B$ of generic entanglement of size factor $n$
against  (a) variance $V=\delta^2$
of the final measurement and (b) variance $V=\Delta^2$ of the measurement reference.
 The dot-dashed line indicates the classical limit, 2.
As the coarsening degree $V$ of the final measurement increases in panel (a),
 the Bell function decreases but this effect can be
compensated by increasing the size $n$ of macroscopic entanglement
(dotted curve: $n = 2$, dashed: $n = 3$, solid: $n = 5$).
However, panel (b) shows that the Bell function rapidly decreases
 independent of $n$ when the measurement reference is coarsened.
}
\label{fig:scheme}
\end{figure}

{\it Bell's inequality with entangled photon number states.-}
We first consider entangled number state of photons,
\begin{equation}
|\psi_n\rangle=\frac{1}{\sqrt{2}}(|n_H\rangle|n_V\rangle+|n_V\rangle|n_H\rangle)
\end{equation}
where $|n_H\rangle\equiv|H\rangle^{\otimes n}$ denotes horizontally polarized $n$ photons and  $|n_V\rangle
\equiv|V\rangle^{\otimes n}$ vertically
polarized. If we set $\left| {{n_H}} \right\rangle  \equiv \left| {{o_n}} \right\rangle ,{\text{ }}\left| {{n_V}} \right\rangle  \equiv \left| {{o_{ - n}}} \right\rangle$, this system is identical to Eq.~(\ref{eq:me}).
We then need to find a physical example of a unitary operation such Eq.~(\ref{unitary}). 
We adopt the unitary operation, 
$U_p(\theta)=\exp[i\theta(|n_H\rangle\langle n_V|+h.c.)]$, 
a rotation about the $x$-axis of the Bloch sphere of a polarized number-state qubit
 $\{|n_H\rangle,|n_V\rangle\}$.  
As this unitary operation depends on the photon number $n$, it needs the
nonlinear Hamiltonian 
$\hat H_n =g( {{\hat a^n_H}}{\hat a_V^{\dag n}}e^{i\phi} + h.c.)$
to be realized.
One can in principle implement this type of highly nonlinear Hamiltonian
by decomposing it into series of Gaussian unitaries and
cubic operations~\cite{Seckin,Petr}. 

\begin{figure}[t]
\centerline{\scalebox{0.32}{\includegraphics{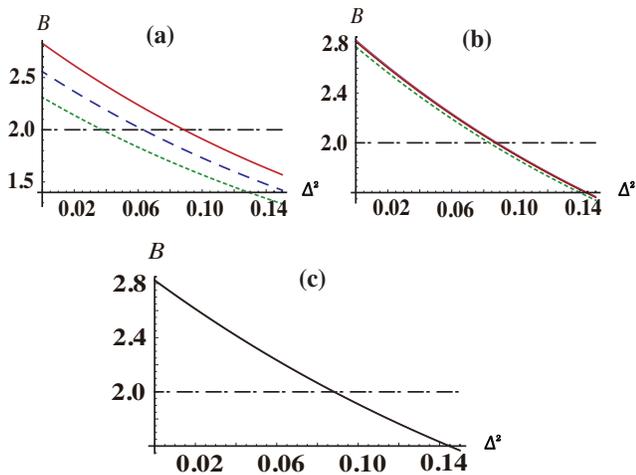}}}
\caption{(color online) Optimized Bell-CHSH function $B$ 
for entangled photon number states with number $n$ against variance $\Delta^2$ of Gaussian coarsening angle for different values of detection efficiency (solid curve: $\eta=1$, 
 dashed: $\eta=0.95$, dotted: $\eta=0.9$) for (a) $n=1$, (b) $n=2$, (c) $n=3$.
In the case of $n=3$, all three curves virtually overlap.
}
\label{resultboth}
\end{figure}

Considering a realistic condition of photon loss, we use a dichotomic measurement operator
\begin{equation}
{\cal O}^p=\sum_{k=1}^n\big(|k_H\rangle\langle k_H|-|k_V\rangle\langle k_V|\big)+|0\rangle\langle 0|
\label{measure}
\end{equation}
and model an inefficient measurement 
using a beam splitter before the final photodetector.
With its efficiency $\eta$, a photodetector  for field mode $a$ is described by a perfect detector after a beam splitter 
 $\hat{\cal B}_{a a'}(\eta)=e^{\zeta({\hat a}^\dagger {\hat a}'
-{\hat a} {\hat a}^{\prime \dagger})/2}$ where $\eta{=}(\cos\zeta)^2$
and $a'$ represents the vacuum mode. The beam splitter parameter $\zeta$, which determines
the transmission ratio $\eta$, represents the degree of coarsening in the final detection.
It was shown \cite{Lim} that  the Bell-CHSH inequality is violated 
even by highly inefficient detectors, {\it i.e.}, $\eta$ very small as $n$ can be made sufficiently large.
In other words,
inefficiency of the final detector can 
be compensated by increasing number $n$ of the entangled photon-number state. 
However, 
when the fuzziness of the unitary operation, $U_p(\theta)$, is considered with a Gaussian noise as in Eq.~(\ref{couni})
without coarsening the final detection,
it is straightforward to show that the correlation function is exactly the same as Eq.~(\ref{unicorrel}).

It would be interesting to consider both the unitary transform and the final detection being coarsened
in order to investigate more realistic scenarios.
The correlation function can be obtained as 
\begin{equation}
{E_{\eta ,\Delta }}({\theta _a},{\theta _b}) = \langle {\cal O}^p_{\eta,\Delta}({\theta _a}) \otimes {\cal O}^p_{\eta,\Delta}({\theta _b}) \rangle_{a,b}
\label{eq:ee}
\end{equation}
where 
\begin{equation}
{\cal O}^p_{\eta,\Delta}(\theta_a)= \int d \theta {P_\Delta }(\theta  - {\theta _a})\left[ {U^\dag(\theta ){\hat{\cal B}_{aa'}^\dagger} {\cal O}^p {\hat{\cal B}_{aa'}} {U }(\theta )} \right]
\label{eq:oo}
\end{equation}
for mode $a$ and ${\cal O}^p_{\eta,\Delta}(\theta_b)$ is likewise defined. 
We then  numerically calculate  the optimized Bell function $B$ for several values of $n$ and plot
the results in Fig.~\ref{resultboth}. It shows that the quantum-to-classical
transition quickly occurs as fuzziness $\Delta$ of the unitary transform increases.
The figure also shows that the decreasing rate of the Bell function caused by
coarsening the unitary transform does not depend on the value of $\eta$.

\begin{figure}[t]
\centerline{\scalebox{0.38}{\includegraphics{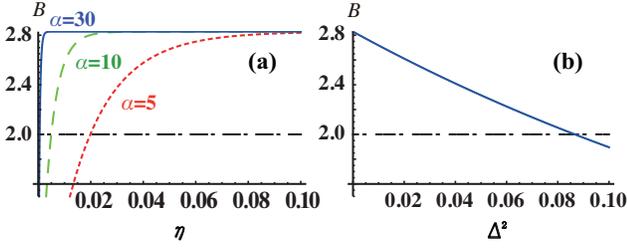}}}
\caption{(color online) Optimized Bell function $B$ for entangled coherent states against (a) homodyne detector efficiency $\eta$ and (b) variance ${\Delta ^2}$ of Gaussian coarsening of 
measurement reference for $\alpha=5$ (dotted curve), $\alpha=10$ (dashed) and $\alpha=30$ (solid).
While the decrease of the measurement efficiency can be made up by increasing $\alpha$ (panel (a)),
coarsening of the measurement reference causes virtually the same decrease of the Bell function regardless of the values of $\alpha$ (panel (b)).}
\label{ecs}
\end{figure}

{\it Bell's inequality with entangled coherent states.-} An entangled coherent state~\cite{Ourj2009,Sanders2012}
$\left| \psi_\alpha \right\rangle  \propto|\alpha ,\alpha \rangle  + | - \alpha , - \alpha \rangle $,
where $|{\pm}\alpha\rangle$ are coherent states of amplitudes $\pm\alpha$, is considered as a macroscopic
quantum state when $\alpha$ becomes large \cite{LeeJeongPRL2011}.
It is known \cite{JeongLeePRA2002} that effective rotations $U_\alpha(\theta)$ in the space spanned by the basis $\{\left| \alpha  \right\rangle, \left| -\alpha  \right\rangle\}$, required for a Bell inequality test, can be performed using single-mode Kerr nonlinearities and displacement operations.
A Bell test can then be  performed using dichotomized homodyne measurements, where an eigenvalue $+1$ ($-1$) is assigned for any positive (negative) outcomes \cite{JeongPRL2009,Lim}.
With homodyne efficiency $\eta$ and Gaussian reference coarsening of $\theta$ with standard deviation $\Delta$,
the correlation function can be obtained in the same way described above using the measurement operator
${\cal O}^h=\int_0^\infty|x\rangle\langle x|dx-\int_{-\infty}^0|x\rangle\langle x|dx$
that replaces
${\cal O}^p$ in Eqs.~(\ref{eq:ee}) and (\ref{eq:oo}).
Our numerical results  \cite{supp} presented in Fig.~\ref{ecs} confirm that 
coarsening of the measurement reference 
cannot be made up by increasing macroscopicity $\alpha$ (Fig.~\ref{ecs}(b)), while it can be made so when the measurement efficiency is coarsened (Fig.~\ref{ecs}(a)). 
Here, we can consider another interesting case where the angle of the homodyne detection,
which should also be controlled precisely as a measurement reference, is coarsened, which leads to qualitatively the same conclusion \cite{supp}.

\begin{figure}[t]
\centerline{\scalebox{0.34}{\includegraphics{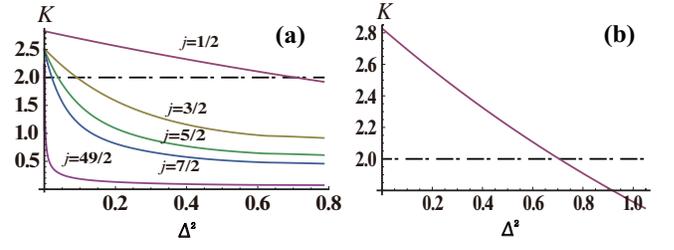}}}
\caption{(color online) (a) Optimized Leggett-Garg function $K$ against variance $\Delta^2$ of Gaussian coarsening 
of the measurement reference for different spin $j$ states. For the larger values of $j$,
the faster decrease of the Leggett-Garg function is observed.
(b) Optimized Leggett-Garg fuction $K$ against variance $\Delta^2$ of Gaussian coarsening time under
 the nonclassical Hamiltonian.
As the temporal reference of the measurement is coarsened by the increase of $V=\Delta^2$, violation of Leggett-Garg inequality disappears regardless of $j$.}
\label{generalj}
\end{figure}

{\it Leggett-Garg inequality with spin systems.-}
The temporal correlation function ${C_{ab}} \equiv \left\langle {Q({t_a})Q({t_b})}
 \right\rangle$ between $t_a$ and $t_b$
for a dichotomic measurement operator $Q$ forms the Leggett-Garg inequality
\begin{equation}
K \equiv {C_{12}} + {C_{23}} + {C_{34}} - {C_{14}} \le 2 
\end{equation}
which is forced by macroscopic realism~\cite{LeggettGarg}.
In the case of the Leggett-Garg inequality that utilizes time sequential measurements,
it is natural to consider coarsening of the temporal references. 	
We shall consider coarsening of two types of unitary operations for spin-$j$ systems 
considered in Refs.~\cite{Brukner07,Brukner08}
with the dichotomized  parity measurement 
$Q = \sum\nolimits_{m =  - j}^j {{{( - 1)}^{j - m}}\left| m \right\rangle \left\langle m \right|}$
where $|m\rangle$ is a spin eigenstate of the spin-$j$ operator ${\hat J}_z$.
The first unitary operation to be considered
is  $U_j(\theta) = {e^{ - i\theta{{\hat J}_x}}}$
with $\theta= \omega t$ and
the initial state is assumed to be the maximally mixed spin-$j$ system
$\sum\limits_{m =  - j}^j 
| m \rangle \langle m |/(2j + 1)$.
We again consider Gaussian coarsening of the unitary operation applied to the measurement operator $Q$
as $Q_{\Delta}(\theta_0)= \int d \theta {P_\Delta }(\theta  - {\theta _0})
\left[ {U_j^\dag(\theta )Q {U_j }(\theta )} \right]$.
The temporal correlation function between $t_a$ and $t_b$ can be obtained as
${C_{ab}} = {p_{{ + _a}{ + _b}}} + {p_{{ - _a}{ - _b}}} - {p_{{ + _a}{ - _b}}} - {p_{{ - _a}{ + _b}}}$,
where ${p_{{ +_a}{ + _b}}}$ is the probability for measuring $+$ at $t_a$ and then $+$ at $t_b$, and so on. 
After some calculation, we obtain 
\begin{equation}
C_{ab}= \sum\limits_{m =  - j}^j {\int_{ - \infty }^\infty  {d\theta '{P_\Delta }(\theta ' - \theta_{b-a} ){e^{2im\theta '}}/\left( {2j + 1} \right)} },
\label{spincorrel}
\end{equation}
where $\theta_{b-a}=\omega(t_b-t_a)$.
We plot the numerically optimized Leggett-Garg function in Fig.~\ref{generalj}(a)
and observe
the decrease of the Leggett-Garg function for any value of $j$  by increasing the coarsening degree of
the measurement reference.
We note that the larger value of $j$ 
leads to more rapid destruction of the Leggett-Garg violation by coarsening the measurement reference.

It was shown \cite{Brukner08} that under a unitary operation that can generate a macroscopic superposition, the Leggett-Garg inequality is violated even with a coarsened measurement.
The corresponding unitary operation 
 is  $U(\theta)=\exp[i\theta(\left| { + j} \right\rangle \left\langle { - j} \right|+h.c.)]$ with $\theta=\omega t$ 
and this is identical to the unitary operation in Eq.~(\ref{unitary}) if
$\left| { \pm j} \right\rangle$ 
 are replaced with $\left| o_{\pm n} \right\rangle$. 
The nonclassical Hamiltonian associated with such a unitary operation is 
\begin{equation}
\hat H = i\omega (\left| { - j} \right\rangle \left\langle { + j} \right| - \left| { + j} \right\rangle \left\langle { - j} \right|).
\label{nonc}
\end{equation}
Assuming an initial state $ \left| { + j} \right\rangle$,
we can calculate the temporal correlation function $C_{ab}$ by the same procedure described above,
and it is found to be
${C_{ab}} = {\text{ }}{{\text{e}}^{ - {\Delta ^2}/2}}\cos [\omega ({t_b} - {t_a})]$
by applying the same Gaussian coarsening of the measurement reference. The temporal correlation function 
$C_{ab}$ is obviously independent of $j$ and
the Leggett-Garg violation disappear by coarsening of the unitary operation 
as plotted in Fig.~\ref{generalj}(b). 
Thus the results for the spin system with the Leggett-Garg inequality are consistent with the previous ones.

{\it Remarks.-}
There have been studies to explain the quantum-to-classical transition: they have focused on
either the evolution of the state or the accuracy of the final measurement resolution.
However, the accuracy of the measurement reference has not been properly considered
in this context. Our study consistently shows that 
when a measurement reference such as the timing and the axis angle of the measurement
is coarsened, it cannot be compensated by increasing ``macroscopicity" of the quantum state
or by using an interaction to generate such macroscopic quantum states. 
This is obviously not the case when only the measurement resolution is coarsened.
Our investigation covers a wide range of physical systems from discrete to continuous
variable systems using various degrees of
freedom such as spins, polarizations, photon numbers and quadrature variables. 
Even though our discussions mainly adopt terminologies in optics, they can be generalized to
various physical systems such as atomic and mechanical systems \cite{supp}.
Our result provides new insight into the quantum-to-classical transition 
from a different angle by revealing the importance of the observer's ability in controlling the measurement reference,
and more generally, importance of preciseness in quantum operations.

This work was supported by the National Research Foundation of Korea (NRF) grant funded by the Korea government (MSIP) (No. 2010-0018295) and by the UK Engineering and Physical Science Research Council.

{\it Note added.-}
At the completion of our work, we became aware of Ref.~\cite{Wang2013}
recently uploaded on a preprint server.
They considered superpositions of coherent states and suggested a conjecture
that outcome precision or control precision has to increase in order to observe quantum effects. 
While the system considered in Ref.~\cite{Wang2013} is different from ours,
the results in their work are consistent with our conclusions
in emphasizing the importance of the ``control precision'' of quantum measurements.

\newpage

~

\newpage

 \renewcommand{\theequation}{A.\arabic{equation}}
  \setcounter{equation}{0}  
\appendix*
\section{Supplementary Material}

\subsection{Testing Bell's inequality with entangled coherent states and coarsening the unitary operations}
One of the most frequently cited non-classical continuous-variable states is the entangled coherent state 
\cite{Sanders2012}
\begin{equation}
\left| {{\psi}} \right\rangle  = {\cal N}(|\alpha ,\alpha \rangle  + | - \alpha , - \alpha \rangle ),
\end{equation}
where $|{\pm}\alpha\rangle$ are coherent states of amplitudes $\pm\alpha$ and 
${\cal N}=\{2(1 + {e^{ - 4{{\left| \alpha  \right|}^2}}})\}^{-1/2}$.
For simplicity, we assume that $\alpha$ is real and positive 
without loss of generality.
Such states are useful for various applications to quantum information processing
\cite{Sanders2012} and were
experimentally realized using optical fields \cite{Ourj2009}.
The generation of the entangled coherent state has been considered in other physical systems including superconducting LC modes \cite{ChePRB2009},  Bose-Einstein condensates \cite{PanPRA2007}, and
motional degrees of freedom of trapped ions \cite{WangSanders2001} and nano-cantilevers/nano-mirrors
\cite{nano1,nano2}. Even though our discussion in this supplementary material uses terminologies in optics, that can
be generalized to other harmonic oscillator systems. 
The degree as a macroscopic quantum state for the entangled coherent state
increase as $\alpha$ becomes larger \cite{LeeJeongPRL2011}.  

The unitary operation $U_\alpha(\theta)$ required to test a Bell-type inequality 
can be performed using single-mode Kerr nonlinearities and displacement operations as \cite{JeongLeePRA2002,SJR,main,Lim}
\begin{equation}
{U_\alpha}(\theta) =  {\hat U_{NL}}\hat D(i\theta /4\alpha ){\hat U_{NL}}
\end{equation}
where $\hat D(\beta ) = \exp [\beta {\hat a^\dag } - {\beta^* }\hat a]$  is
the displacement operator and ${\hat U_{NL}} = \exp [ - i\pi {({\hat a^\dag }\hat a)^2}/2]$ represents
a single-mode Kerr nonlinear interaction.
It is known \cite{JeongLeePRA2002,SJR,main} that the operation $U_\alpha(\theta)$ well approximates perfect $x$-rotations in the vector space spanned by the basis $\{\left| \alpha \right\rangle, \left| -\alpha  \right\rangle\}$ when $\alpha$ is not too small.
A Bell inequality test can be performed using a dichotomized quadrature measurement
\begin{equation}
{{\cal O}^h} = \int_0^\infty  {\left| x \right\rangle } \left\langle x \right| - \int_{ - \infty }^0 {\left| x \right\rangle } \left\langle x \right|
\end{equation}
where $|x\rangle$ is an eigenstate of a quadrature measurement outcome $x$.
It is well known that homodyne measurements of optical fields realize such measurements \cite{QOtext}.

We first consider the case of an inefficient homodyne measurement with efficiency $\eta$
that can be modeled by a beam splitter with transmission ratio $\eta$ right before 
a perfect homodyne detector.
The correlation function can be obtained as
\begin{equation}
{E_\eta }({\theta _a},{\theta _b}) =  \langle{\cal O}_\eta ^h({\theta _a}) \otimes {\cal O}_\eta ^h({\theta _b})\rangle _{a,b}
\label{eq:aver}
\end{equation}
with 
\begin{equation}
{\cal O}_\eta ^h(\theta_a) = U_\alpha^\dag ({\theta_a })\hat {\cal B}_{aa'}^\dag {{\cal O}^h}\hat {\cal B}_{aa'}^{}U_\alpha^{}({\theta_a }),
\end{equation}
where $\hat{\cal B}_{aa'}$ is the beam splitter operator with the transmission ratio $\eta$ defined in the main Letter, $a'$ is the vacuum mode, and ${\cal O}_\eta ^h(\theta_b) $ for mode $b$ is defined likewise.
It should be noted that vacuum modes $a'$ and $b'$ employed to implement the beam splitter operators
should be traced out in order to finally obtain the average value in Eq.~(\ref{eq:aver}).
An explicit form of the correlation function is
\begin{equation}
{E_\eta }({\theta _a},{\theta _b}) = \frac{1}{{1 + {e^{ - 4{\alpha ^2}}}}}{[{\text{erf}}(\sqrt {2\eta } \alpha )]^2}\cos [2({\theta _a} - {\theta _b})].
\label{result-a}
\end{equation}

\begin{figure}[t]
\centerline{\scalebox{0.4}{\includegraphics{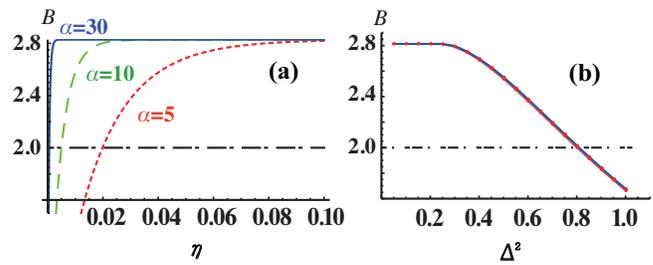}}}
\caption{
Optimized Bell function $B$ against (a) homodyne detector efficiency $\eta$ and (b) variance ${\Delta ^2}$ 
of Gaussian coarsening of the homodyne angle $\lambda$ with $\alpha=5$ (dotted curve), $\alpha=10$ (dashed) and
$\alpha=30$ (solid). In panel (a), inefficiency of the homodyne detection is compensated by increasing
the value of $\alpha$ while it is not the case with panel (b).
} 
\label{ecss}
\end{figure}

We now move to coarsening of the measurement reference.
First, the parameter  $\theta$ of the unitary transform $U_\alpha(\theta)$ may be coarsened with
Gaussian standard deviation $\Delta$. Under the assumption of the perfect measurement efficiency ({\it i.e.}, $\eta=1$), the measurement operator is
\begin{equation}
{\cal O}_\Delta ^h({\theta _0}) = \int {d\theta } {P_\Delta }(\theta  - {\theta _0})[U_\alpha^\dag (\theta ){{\cal O}^h}U_\alpha (\theta )]
\end{equation}
where ${P_\Delta }(\theta  - {\theta _0})$ is the Gaussian kernel centered at $\theta _0$ with standard deviation $\Delta$ and the 
correlation function is 
\begin{equation}
{E_\Delta }({\theta _a},{\theta _b}) = {\langle O_\Delta ^h({\theta _a}) \otimes O_\Delta ^h({\theta _b})\rangle _{a,b}}.
\end{equation}
We obtain the correlation function as
\begin{equation}
{E_\Delta }({\theta _a},{\theta _b}) = \frac{{{e^{ - 4{\Delta ^2}}}}}{{1 + {e^{ - 4{\alpha ^2}}}}}{{\text{[erf(}}\sqrt 2 \alpha )]^2}\cos [2({\theta _a} - {\theta _b})].
\label{result-b}
\end{equation}
Equations~(\ref{result-a}) and (\ref{result-b}) can be used to construct Bell functions and
numerically optimized results are plotted for several values of $\alpha$ in our main Letter.

Another interesting case to be considered is to coarsen the angle of the homodyne detection as
a measurement reference.
The quadrature variable to be measured is
\begin{equation}
\hat{x}_\lambda=\frac{1}{\sqrt{2}}(\hat a e^{-i\lambda}+\hat a^\dagger e^{i\lambda})
\end{equation}
where $\lambda$ is the phase angle to the {\it reference} quadrature $x$.
Ideally, this angle  should be maintained as $\lambda=0$ during the measurement process,
while we assume that  it is not accurately controlled under the Gaussian coarsening of standard
deviation $\Delta$.
Since the measurement efficiency $\eta$ is supposed to be perfect, the correlation function can be
obtained in the same way explained above using $|x_\lambda\rangle$ instead of $|x\rangle$ 
and $P_\Delta(\lambda-\lambda_0)$ instead of $P_\Delta(\theta-\theta_0)$.
The correlation function in this case is obtained as
\begin{widetext}
\begin{equation}
E_{\Delta}(\theta_a,\theta_b) = \frac{1}{{1 + {e^{ - 4{\alpha ^2}}}}}{\left( {\int_{ - \infty }^\infty {\frac{{{e^{ - \tfrac{{{\lambda ^2}}}{{2{\Delta ^2}}}}}\alpha \cos (\lambda ){\text{erf[}}\sqrt {{\alpha ^2}(1 + \cos (2\lambda ))} ]}}{{\Delta \sqrt {2\pi {\alpha ^2}{{\cos }^2}(\lambda )} }}d\lambda } } \right)^2}
 \times \cos [2({\theta _a} - {\theta _b})].
\label{result2}
\end{equation}
\end{widetext}
We obtain and plot the numerically optimized Bell function for several cases of $\alpha$ in Fig.~\ref{ecss}
together with the case of the measurement inefficiency based on Eq.~(\ref{result-a}) 
for a comparison. It is obvious that the inefficiency of the  homodyne measurement  can be
compensated by increasing the macroscopicity of the entangled coherent state (Fig.~\ref{ecss}(a))
while the coarsening of the measurement reference (homodyne angle) forces the quantum-to-classical transition
(Fig.~\ref{ecss}(b)).

\end{document}